\documentstyle[sprocl]{article}

\def\be{\begin{equation}}
\def\ee{\end{equation}}
\def\bea{\begin{eqnarray}}
\def\eea{\end{eqnarray}}

\begin{document}

\title{TOWARDS EVEN AND ODD SQUEEZED NUMBER STATES}

\author{ MICHAEL MARTIN NIETO\footnote
{Email: mmn@pion.lanl.gov } }

\address{Theoretical Division (T-8, MS-B285), 
Los Alamos National Laboratory, \\
University of California,
Los Alamos, New Mexico 87545 U.S.A.\\
and\\
 Universit\"at Ulm, 
Abteilung f\"ur Quantenphysik, \\
Albert-Einstein-Allee 11, 
D 89069 Ulm, GERMANY}


\maketitle\abstracts{
The time evolution of even and odd squeezed states, as well as 
that of squeezed number states, has been given in simple, analytic 
form.  This follows experimental work on trapped ions which has 
demonstrated even and odd coherent states, number states, and 
squeezed (but not displaced) ground states.  We review this situation
and consider the extension to even and odd squeezed number states.
Questions of uncertainty relations are also discussed.}
  
In 1926, looking for quantum states which could follow the classical
motion, Schr\"odinger discovered what we now call ``coherent states" (CS).   
He did this before Heisenberg 
discovered his uncertainty relation (HUR).  
Even so, Schr\"odinger's logic most closely followed that of 
the minimum-uncertainty (MU) method 
for obtaining CS we discuss below.  
Having the work of Schr\"o-dinger and Heisenberg as a background, Kennard
discovered what we now call ``squeezed states" (SS) in 
1927.\cite{ssur} 

A few years later, Schr\"odinger synthesized these
concepts in devising his own uncertainty relation (SUR).\cite{SUR}  
Starting from two non-commuting operators, 
$[X,P]=iG$, Schr\"odinger showed that the uncertainty product 
satisfies
\begin{equation}
(\Delta X)^2(\Delta P)^2 \ge
|\langle  [X,P]/2 + 
\{\bar{X},\bar{P}\}/2 \rangle|^2 
\ge  \langle G \rangle^2/4.   
\end{equation}
The first inequality is derived from the Schwartz inequality.  It holds if  
\begin{equation}
\bar{X} |\psi\rangle \equiv (X-\langle X\rangle) |\psi\rangle 
= \lambda (P-\langle P\rangle) |\psi\rangle ,  \label{sureq}
\end{equation}
and is the SUR.
The states which satisfy this, for the harmonic oscillator (HO),
where $X=x$ and $P=p$, are 
the squeezed-state Gaussians.  The second inequality holds, 
yielding the HUR, when the expectation
value of the anticommutator is zero: $ 0 = 
\langle \frac{1}{2}\{\bar{X},\bar{P}\}\rangle$.  For the HO, 
this means the width of the Gaussians are that of the ground state,
and one has the CS.  

The above is the minimum-uncertainty (MU) method.  But by comparing 
Eq. (\ref{sureq}) with the defining equations for ladder-operator
(LO) CS, $ 
a|\alpha\rangle = \alpha|\alpha\rangle$,  
and SS, $
   [ \mu a - \nu a^{\dagger} ] 
|\alpha, z \rangle =
   \beta(\alpha,z)|\alpha,z \rangle$,
one sees that they are equivalent.

The displacement-operator (DO) CS and SS, are defined as
$
|\alpha\rangle = D(\alpha)|0\rangle$, where 
$D(\alpha)=\exp[\alpha a^{\dagger}-\alpha^* a]$, and 
$|\alpha,z\rangle = D(\alpha)S(z)|0\rangle$,
where $
S(z)=\exp[\frac{1}{2}za^{\dagger}a^{\dagger}-\frac{1}{2}z^* aa]$,
respectively.
That they are equivalent to the MU and LO states follows from the commutators
of $a$ with $D$ and $S$.  

All these methods have been extended to many potential and group symmetry 
systems.  We are concerned with generalizations to other states in the 
HO potential, because they are now 
being studied using ions in a trap.


Before continuing, I want to emphasize  two meanings of CS and SS 
that are in use.  To a mathematical physicist, it 
is the precise mathematical definitions I have used and their
generalizations to other systems.  They specifically search for  
wave packets that best follow 
the classical motion and maintain their shapes (CS)
and wave packets that best follow the classical motion while having 
the uncertainty oscillate between the $X$ and $P$ operators (SS).  
An experimentalist is interested in states which have these properties,
precisely or not.
If the uncertainties are oscillating between the quadratures, 
there is a squeezing.  It is fair enough for him to call these 
``squeezed states."


Now consider 
the even and odd (EO) coherent\cite{manko} and squeezed\cite{eoss} 
states.  They can be obtained, from 
the MU and LO points of view, as the solutions to  equations of the forms
\begin{equation}
aa|\alpha\rangle_{\pm} = \alpha^2|\alpha\rangle_{\pm},  
~~~~~~~~~~~~~
   [ \mu aa - \nu a^{\dagger} a^{\dagger}] 
|\alpha, z \rangle_{\pm} =
   \beta(\alpha,z)|\alpha,z \rangle_{\pm}.  \label{eocs}
\end{equation}
Even starting from the Fock representation, the solution 
of Eq. (\ref{eocs}a) leads to wave functions that are parity even and odd
reflected sums of two displaced ground-state Gaussians.  The 
wave function solutions to Eq. (\ref{eocs}b) are, on the face of it,  
complicated confluent hypergeometric functions.  

\indent From the DO point of view,  one can 
mathematically create the above EOCS by applying
$
D_{\pm}|0\rangle \equiv N_{\pm}[D(\alpha)\pm D(-\alpha)]|0\rangle
        = |\alpha\rangle_{\pm}
$
where $D(\alpha)$
is the ordinary displacement operator. 
But even forgetting that one has  
to put the normalization in by hand, 
there is, in principle, a mathematical 
problem.   The ``displacement operator" $D_{\pm}(\alpha)$
is NOT unitary, 
$D_{\pm}^{\dagger}(\alpha)D_{\pm}(\alpha) \ne 1$.  

Coming to the DO EOSS, there is another problem.  For
the ordinary squeezed states, the squeeze operator
$S(z)=\exp[(za^{\dagger}a^{\dagger}-z^*aa)/2]$ changes $S^{\dagger}aS$
into a combination of $a$ and $a^{\dagger}$.  
But there is no squeeze operator $\hat{S}$  that changes $aa$ 
into a sum of $aa$ and $a^{\dagger}a^{\dagger}$.  

Even so, one can  blindly proceed,  and ask for the form of 
$D_{\pm}(\alpha)S(z)|0\rangle$.   One obtains even and odd parity 
squeezed Gaussians. Then applying the 
harmonic-oscillator time evolution operator, 
$T(t)=\exp[-it(a^{\dagger}a+1/2)]$, this
leads to  analytic wave functions and analytic probability 
densities.  Taking, for simplicity, $z=re^{i\phi}$ to be real, 
the densities are
\begin{equation}
  \rho_{s\pm}(x,t) = N_{\pm}(t)  
       \left\{\cosh\left[{2xx_0(\cos t)}{/d^2}\right]
            \pm \cos\left[{2xx_0 (\sin t)}{/(d^2 s^2)}\right]\right\},
\end{equation}
where $\ln s = r~sgn(r) $ and $d^2 = s^2 \cos^2t + \sin^2t/s^2 $.
Elsewhere,\cite{eoss}  graphs are given of the time evolution of such 
wave packets, showing the even and odd  
interference patterns near the origin and the evolutions of the squeezing.

These same techniques were then used to analyze the time evolution of 
squeezed number states: $T(t)D(\alpha)S(z)|n\rangle$.\cite{sns}
The wave functions are similar to ordinary squeezed states, 
except they are multiplied by a Hermite polynomial $H_n(Y)$, 
where $Y$ is a squeezed and displaced, time-evolving position.  
In particular, three dimensional-plots of the 
the probability density for the $n=1$ case, $\rho_{s(1)}(x,t)$,  
show the time-evolution 
for various squeezings.\cite{sns}  The figures are 
similar to those showing the time-evolution of squeezed states.  
But the center of the packet has not one but two humps.

To calculate the time-evolution of squeezed even and odd number 
states, one can combine all the above techniques.  
In particular, one wants to find
$
T(t)D_{\pm}(\alpha)S(z)|n\rangle \rightarrow \Psi_{s(n)\pm}(x,t)$
and $\rho_{s(n)\pm}(x,t) =
           \Psi^*_{s(n)\pm}(x,t)\Psi_{s(n)\pm}(x,t)$.
This program is now underway.  The results will  combine
the previous two findings.  There will be even and odd 
interferences when the two sides of the wave packet come together 
at the origin, but there will be the added complication of each 
half of the wave packet having $n+1$ humps, instead one one.  

There is hope that
all the states discussed here can be observed.  This hope is based
on the recent work of Wineland's group with trapped 
$Be^+$ ions.\cite{wine}  By entangling the electronic 
and motional states they produced  even and odd coherent (displaced)
states.  They also have been able to produce 
number states and squeezed 
(but not displaced) ground states.  If all these
techniques can be combined then in principle 
even and odd squeezed states, displaced and squeezed number
states, and  perhaps even even and odd squeezed number states 
could be produced.  However, to do so will be a real 
accomplishment.

This work was supported by the U.S. Department of Energy and the 
Alexander von Humboldt Foundation.

\end{document}